\documentclass[5p,times]{elsarticle}

\usepackage{amsmath,amssymb}
\usepackage{booktabs}
\usepackage{graphicx}
\usepackage{microtype}
\usepackage{hyperref}
\usepackage{enumitem}

\hypersetup{hidelinks}
\graphicspath{{figures/}{./}}

\journal{Simulation Modelling Practice and Theory}

\begin{document}

\begin{frontmatter}

\title{Ground-Truth-Aware Stress Testing of a Closed-Loop Smart-Building Digital Twin Under Sensor Drift and Missing Data}

\author{Lyes Saad Saoud}

\begin{abstract}
Digital twins are increasingly used to support monitoring and control in smart buildings, yet many evaluations emphasize state-estimation accuracy or fault detection rather than the downstream question that ultimately matters for closed-loop operation: \emph{when do sensing errors materially change control outcomes?} This paper introduces a ground-truth-aware simulation framework that separates a latent physical state from a corrupted sensing layer and compares practical sensor-driven policies against an oracle controller with direct access to the latent state. The synthetic smart-building twin contains 20 zones observed at 15-minute intervals over 30 days and models occupancy-driven CO$_2$, ventilation--energy trade-offs, additive sensor drift, measurement noise, and missing observations. Closed-loop decisions use a one-step delay, and policy comparisons employ common random numbers for paired Monte Carlo evaluation. Under nominal sensing, a raw-sensor threshold controller disagreed with the oracle on 2.59\% of decision steps, yet its CO$_2$ exceedance gap was $-1.25\times10^{-4}$ with a 95\% Monte Carlo interval of $[-4.65\times10^{-3},4.69\times10^{-3}]$, while mean energy and comfort were effectively unchanged. At $8\times$ nominal drift, raw-sensor decision mismatch increased to 7.00\%, but the corresponding CO$_2$ exceedance, energy, and comfort gaps remained small. Across a $4\times4$ drift--missingness stress grid, none of the 48 evaluated sensor-driven policy--condition combinations crossed the predeclared material-divergence thresholds. An ablation study further showed that a three-sample rolling median increased nominal mismatch from 2.59\% to 6.31\% without a meaningful outcome advantage; the full fault-aware policy inherited this behavior. Finally, scalar Ground-Truth Regret rankings changed with utility weights, demonstrating that a single policy ranking is objective-dependent. The results establish a useful distinction between \emph{decision disagreement} and \emph{outcome degradation}: sensor corruption can change individual control actions without necessarily producing materially different aggregate physical outcomes. Because all experiments are synthetic and uncalibrated to a real building, the contribution is methodological rather than a claim of real-world building performance.
\end{abstract}

\begin{keyword}
digital twin \sep smart buildings \sep sensor drift \sep missing data \sep closed-loop control \sep robustness \sep Monte Carlo simulation \sep verification
\end{keyword}

\end{frontmatter}

\section{Introduction}
\label{sec:introduction}
\begin{figure*}[t]
    \centering
    \includegraphics[width=\textwidth]{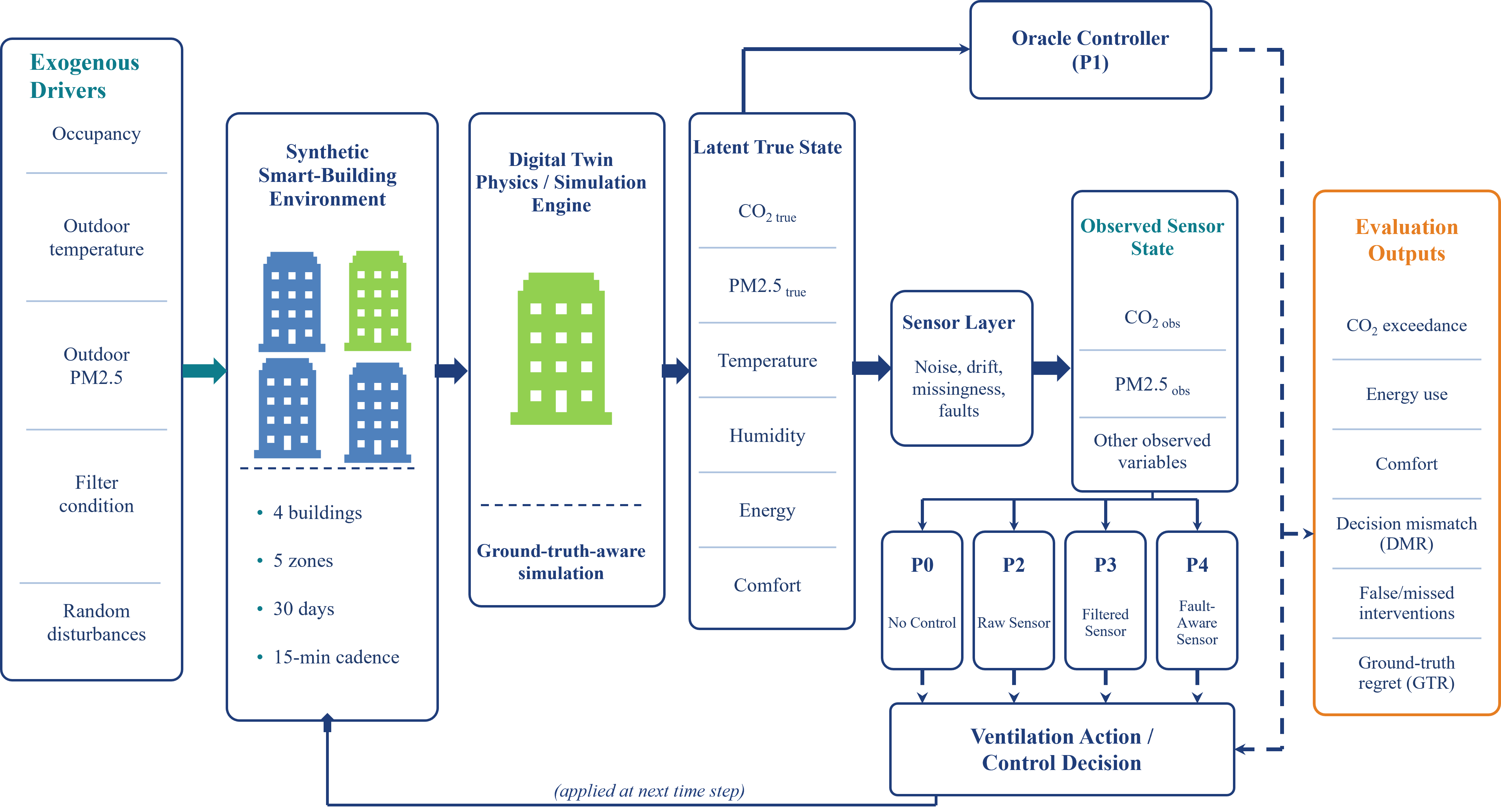}
    \caption{Ground-truth-aware closed-loop digital-twin testbed. Exogenous drivers and the physics-based simulator generate the latent state; the sensor layer produces corrupted observations for the practical controllers, while the oracle (P1) reads the latent state directly. Ventilation actions are applied at the next 15-min step, and policies are evaluated using decision- and outcome-level metrics under sensing stress. The initial schematic was prepared using the OpenAI ChatGPT image-generation service (August 2026) and subsequently reviewed and edited by the author.}
    \label{fig:architecture}
\end{figure*}
Digital twins combine virtual models, data, and computational services to represent and reason about physical systems. Their scope ranges from virtual representations and synchronization mechanisms to prediction, optimization, and closed-loop decision support \cite{kritzinger2018categorical,rasheed2020values,fuller2020enabling,jones2020characterising}. In the built environment, digital-twin research has expanded from conceptual frameworks to operational use cases involving monitoring, energy management, and control \cite{khajavi2019buildings,boje2020semantic,delgado2021built,ohueri2024decarbonizing}. Recent work has also emphasized synchronization between physical and virtual states in building control loops \cite{koo2026synchronization}.

A central difficulty is that the digital twin does not observe the physical state directly. Building control systems depend on sensor measurements that may be noisy, biased, drifting, missing, miscalibrated, or otherwise faulty. This problem is well established in building fault detection and diagnostics (FDD) \cite{katipamula2005review,kim2018review,zhang2021sensor}, in general intelligent sensing \cite{erhan2021anomaly}, and in recent smart-building sensor-fault studies \cite{chahine2026brick}. Public HVAC fault datasets have helped make fault-detection methods reproducible \cite{granderson2023dataset}, but a different question remains underexplored: \emph{how much do sensing errors matter to the decisions and physical outcomes of a closed-loop digital twin?}

That question is not equivalent to fault-detection accuracy. A corrupted sensor may cause the controller to choose a different action than an ideal oracle, yet the physical outcome may remain nearly unchanged. Conversely, a small measurement error near a control threshold can alter an action with disproportionate consequences. Evaluating sensor quality only at the measurement layer can therefore miss the decision-level and outcome-level significance of sensor corruption.

This paper develops a fully synthetic, ground-truth-aware closed-loop testbed for studying this distinction. As illustrated in Fig.~\ref{fig:architecture}, exogenous building and environmental drivers are propagated through a physics-based simulation engine to generate a latent physical state. A separate sensor layer converts that latent state into imperfect observations through noise, drift, missingness, and sensor faults. The observed state is supplied to the practical control policies, whereas the oracle controller has access to the latent ground-truth state. Ventilation decisions are applied at the next time step, preserving the causal structure of the closed loop. This separation between latent state, observed state, decision, action, and subsequent physical response makes it possible to determine whether sensor corruption merely changes controller decisions or materially changes building outcomes.

The twin is deliberately not calibrated to a real building; it is a controlled experimental environment for asking whether, and under what stress conditions, observation errors propagate into consequential control outcomes. Following established simulation practice, verification is separated from real-world validation \cite{sargent2010vv,law2015simulation}.

The main contributions are:
\begin{enumerate}[leftmargin=*]
    \item a ground-truth-aware closed-loop architecture that separates latent state, sensor corruption, observed state, control decision, ventilation action, and physical outcome while enforcing a one-step information delay;
    \item a paired Monte Carlo evaluation framework using common random numbers (CRN) to compare oracle, raw-sensor, filtered, fault-aware, and no-control policies under matched exogenous conditions;
    \item decision-level metrics---Decision Mismatch Rate (DMR), False Intervention Rate (FIR), and Missed Intervention Rate (MIR)---reported jointly with physical outcome gaps in CO$_2$, energy, and comfort;
    \item stress tests over sensor drift, missing observations, combined drift--missingness conditions, and an unseen operating distribution, plus an ablation study of the fault-aware controller; and
    \item the empirical finding that, for the evaluated threshold-control dynamics, sensor errors can measurably change decisions while aggregate physical outcomes remain close to the oracle, and that added preprocessing complexity can worsen decision agreement without improving outcomes.
\end{enumerate}

\section{Related Work}
\label{sec:related_work}

\subsection{Digital twins for buildings and closed-loop operation}
Definitions of digital twins differ across application domains, but recurring elements include a physical entity, virtual representation, data connection, synchronization, and computational processes for prediction or decision support \cite{kritzinger2018categorical,jones2020characterising}. Modeling-oriented perspectives emphasize that a twin's value depends on the fidelity and uncertainty of the models and data that connect the physical and virtual systems \cite{rasheed2020values}. For buildings, Khajavi et al. \cite{khajavi2019buildings} outlined the benefits and boundaries of digital twins, while Boje et al. \cite{boje2020semantic} and Davila Delgado and Oyedele \cite{delgado2021built} connected digital-twin concepts to the built environment and construction/operation processes. Reviews of operating-building twins identify energy and decarbonization as prominent use cases \cite{ohueri2024decarbonizing}.

The importance of synchronization becomes sharper once a twin participates in control. Koo and Yoon \cite{koo2026synchronization} recently demonstrated a bidirectional synchronization method in an operational HVAC control loop and showed that improved synchronization can reduce digital-twin model error. Our work addresses a complementary question. Rather than improving synchronization fidelity, we intentionally degrade the observation layer and quantify how far sensor-driven decisions and outcomes move from an oracle reference.

\subsection{Sensor faults and building FDD}
Sensor degradation has long been recognized as a limiting factor in building automation. Reviews by Katipamula and Brambley \cite{katipamula2005review} and Kim and Katipamula \cite{kim2018review} organize model-based, rule-based, and data-driven approaches to building FDD. Zhang et al. \cite{zhang2021sensor} specifically review sensor impact and verification for building-energy FDD and note that calibration, placement, maintenance, and sensor quality directly influence downstream diagnostic performance. Broader anomaly-detection literature similarly treats sensor drift, data quality, and deployment constraints as core challenges \cite{erhan2021anomaly}.

Recent work continues to develop fault-aware sensing for buildings. Chahine and Noura \cite{chahine2026brick} study virtual temperature sensors for detecting bias faults in closed-loop HVAC systems, illustrating how a control loop can mask or redistribute fault signatures. Granderson et al. \cite{granderson2023dataset} provide a large labeled HVAC dataset spanning faulted and fault-free states, enabling reproducible research on such methods. These studies motivate robust sensing and diagnosis, but they do not eliminate the need to ask whether a measurement fault is actually consequential for a specific controller and objective.

\subsection{Gap: from sensor error to outcome error}
The literature therefore contains substantial work on digital-twin fidelity, building FDD, sensor anomalies, and energy-oriented data fusion \cite{himeur2020fusion}. A particularly close recent benchmark is SmartBuildSim \cite{miller2025smartbuildsim}, which generates reproducible synthetic smart-building streams with configurable noise, drift, and missingness for AI benchmarking. Its focus is synthetic data generation and algorithm benchmarking rather than paired oracle-relative propagation of sensing errors from control decisions to simulated physical outcomes. Complementary recent building studies address bidirectional digital-twin synchronization in operational HVAC control \cite{koo2026synchronization} and sensor-bias detection and isolation in closed-loop HVAC systems \cite{chahine2026brick}. The present work instead isolates the chain from sensing corruption to action disagreement and then to physical outcome gaps through paired, ground-truth-aware closed-loop experiments.

What remains less explicit is a three-layer separation between (i) sensor error, (ii) decision disagreement, and (iii) physical outcome degradation. The present study is designed around that separation. By retaining the latent state, the simulator provides a counterfactual reference controller, enabling determination of whether sensor-induced decision differences are consequential or benign within the declared model.

\section{Ground-Truth-Aware Closed-Loop Twin}
\label{sec:closed_loop_twin}

\subsection{Synthetic Research Environment}
The base synthetic environment contains four buildings, five zones per building, 30 days, and a 15-min sampling interval. Each Monte Carlo run therefore contains 57,600 zone-timestep observations. The base random seed is 2026. Occupancy follows a working-hours profile with lower activity during nights and weekends, while outdoor temperature contains both diurnal variation and gradual 30-day variation. Additional simulated variables include humidity, outdoor and indoor PM$_{2.5}$, filter efficiency, noise, energy, and comfort. The closed-loop experiments in this paper, however, use CO$_2$ as the controlled quantity and do not make claims about the realism of the auxiliary PM$_{2.5}$ model.

The simulation is fully synthetic. Parameter values are declared modeling choices and are not estimates from Chicago State University or any other real facility. This distinction is essential: the study supports verification and controlled stress testing rather than real-building validation. Figure~\ref{fig:architecture} summarizes the overall architecture; the following subsections specify its latent dynamics, sensor model, timing, and policies.

\subsection{Latent CO$_2$ Dynamics}
Following the standard single-zone, well-mixed CO$_2$ mass-balance relation \cite{emmerich2001co2dcv,persily2015co2standards}, let $N_t$ denote occupancy, $v_t$ the ventilation rate, $C_{\mathrm{out}}$ the outdoor CO$_2$ concentration, $e$ an occupancy-emission coefficient, and $V$ the zone-volume scaling term. With occupant generation represented by $G_t=eN_t$ and effective ventilation flow represented by $Q_t=v_tV$ in internally consistent model units, the simulator uses the instantaneous equilibrium target
\begin{equation}
C_t^{\star}=C_{\mathrm{out}}+\frac{eN_t}{v_tV}.
\label{eq:target}
\end{equation}

The transient solution of a single-zone mass balance approaches equilibrium exponentially \cite{emmerich2001co2dcv}. Motivated by that structure, the simulator uses the following discrete first-order relaxation with a fixed, declared time constant rather than a calibrated ventilation-dependent time constant:
\begin{equation}
\bar{C}_t
=
\bar{C}_{t-1}
+
\alpha\left(C_t^{\star}-\bar{C}_{t-1}\right),
\qquad
\alpha=1-\exp\!\left(-\frac{\Delta t}{\tau}\right),
\label{eq:dynamics}
\end{equation}
with $\Delta t=15$ min and $\tau=45$ min. Process noise $\epsilon_t$ is then added to obtain the latent state
\begin{equation}
C_t=\bar{C}_t+\epsilon_t.
\label{eq:latent}
\end{equation}

The nominal ventilation baseline is 2.2 air changes per hour (ACH), and an active intervention multiplies the baseline ventilation by 1.8, subject to configured ventilation limits. The 1000 ppm controller threshold is a declared experimental setpoint rather than a universal health or indoor-air-quality limit. NIST notes that approximately 1000 ppm has often been used as a de facto ventilation marker, while cautioning that its meaning depends on the underlying ventilation and occupancy context \cite{persily2015co2standards}.

Energy per interval is modeled as the sum of a base load, a ventilation term, an outdoor-temperature deviation term, an occupancy plug-load term, and additive noise. Comfort is a bounded score in $[0,100]$ that penalizes CO$_2$ above 800 ppm, indoor-temperature deviation from the setpoint, noise above 45 dB, and humidity deviation from 45\%. The 800 ppm comfort breakpoint and the remaining score parameters are synthetic shaping choices, not regulatory thresholds. These equations are not asserted as calibrated building physics; they provide transparent and testable mechanisms for the closed-loop experiment.

\subsection{Sensor Layer}
The observed sensor value is generated from the latent CO$_2$ state as
\begin{equation}
y_t=\operatorname{clip}\!\left(C_t+d_t+\eta_t,\;380,\;5000\right),
\label{eq:obs}
\end{equation}
where $d_t$ is a cumulative random-walk drift process and $\eta_t$ is measurement noise. Under explicit missingness stress, $y_t$ is replaced by NaN with the specified probability. Missing values are never silently converted to zero.

Nominal drift uses a standard deviation of 6 ppm/day before severity scaling. Drift experiments multiply this nominal level by $\{0,0.5,1,2,4,8\}$. Missingness experiments use effective rates $\{0,0.01,0.05,0.10,0.20\}$. The clipping range, drift magnitude, and missingness levels are declared experimental design choices rather than calibrated sensor specifications.

\subsection{One-Step Information Timing}
The timing represented in Fig.~\ref{fig:architecture} prevents circular logic. At time $t$, the controller observes only information available through $t-1$, chooses the ventilation action for interval $t$, and the simulator then evolves the latent state under that action and the current exogenous conditions. The sensor layer subsequently produces the observation that becomes available for the next decision. Thus no policy can use future or same-step outcome information. The same delay is enforced for the oracle; its advantage is access to the latent state rather than access to future information.

\subsection{Policies}
Five policies are evaluated (Table~\ref{tab:policies}). The oracle is a reference rather than a deployable controller.

\begin{table}[t]
\caption{Closed-loop policies.}
\label{tab:policies}
\centering
\small
\begin{tabular}{p{0.16\linewidth}p{0.74\linewidth}}
\toprule
Policy & Decision rule at time $t$ \\
\midrule
P0 No control & Never activates the ventilation boost. \\
P1 Oracle & Activates if latent $C_{t-1}\geq1000$ ppm. \\
P2 Raw & Uses the latest valid observed CO$_2$; missing data hold the previous action. \\
P3 Filtered & Uses the rolling median of the previous three valid observations; missing data hold the previous action. \\
P4 Fault-aware & Adds rolling median, jump/range rejection, and occupancy fallback when the CO$_2$ estimate is unavailable. \\
\bottomrule
\end{tabular}
\end{table}

The fault-aware controller is intentionally simple and interpretable. It rejects recent observations that fall outside the configured 380--5000 ppm range or jump by more than 800 ppm relative to the previous valid reading. If an estimate is unavailable, it falls back to an occupancy trigger at 50\% of zone capacity.

\section{Evaluation Methodology}
\label{sec:evaluation}

\begin{figure*}[t]
\centering
\includegraphics[width=0.98\textwidth]{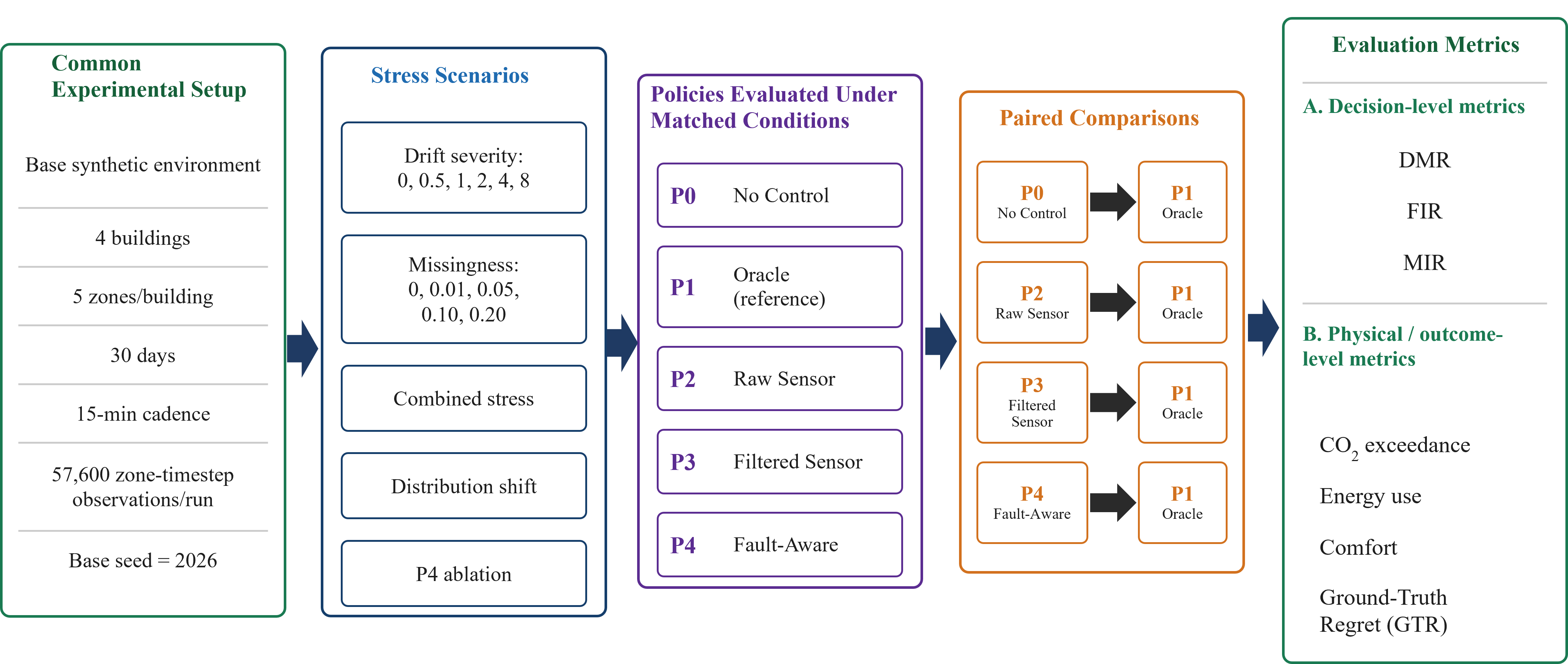}
\caption{Paired Monte Carlo evaluation framework. Within each replicate, all policies share the same exogenous realization and stochastic conditions through common random numbers (CRN). Policies are evaluated under drift, missingness, combined stress, distribution shift, and fault-aware ablation, with decision-level and physical-outcome metrics reported relative to the oracle reference. The initial schematic was prepared using the OpenAI ChatGPT image-generation service (August 2026) and subsequently reviewed and edited by the author.}
\label{fig:evaluation_framework}
\end{figure*}

Figure~\ref{fig:evaluation_framework} summarizes the paired evaluation design. Each Monte Carlo replicate evaluates all policies under matched exogenous and stochastic conditions so that policy differences are not confounded by different sampled scenarios.

\subsection{Common Random Numbers and Monte Carlo Design}
For each Monte Carlo seed, all policies share identical occupancy, outdoor temperature, outdoor PM$_{2.5}$, filter trajectory, baseline ventilation, physical disturbances, drift draws, and measurement-noise draws. Only the control decisions differ. This common-random-number design reduces extraneous variance in paired policy comparisons and follows standard simulation methodology \cite{law2015simulation}.

The full experiments use 100 runs for nominal evaluation, 50 runs at each single-axis drift and missingness level, 30 runs per cell in the two-dimensional stress grid, 100 runs for the distribution-shift comparison, and 100/50 runs for nominal/stress ablation. The complete experimental suite required approximately 168 s in the supplied execution environment (Python 3.12.3). Eleven automated verification tests passed, including no-future-information checks, CRN identity across policies, explicit NaN handling, oracle/raw information separation, metric sign checks, and preservation of the original 57,600-row base simulator.

\subsection{Decision Metrics}
Let $a_t^{p}\in\{0,1\}$ be the action of policy $p$ and $a_t^{o}$ the oracle action. Over $T$ decision points,
\begin{align}
\mathrm{DMR}_p &= \frac{1}{T}\sum_t \mathbb{1}[a_t^p\ne a_t^o],\\
\mathrm{FIR}_p &= \frac{1}{T}\sum_t \mathbb{1}[a_t^p=1,a_t^o=0],\\
\mathrm{MIR}_p &= \frac{1}{T}\sum_t \mathbb{1}[a_t^p=0,a_t^o=1].
\end{align}
DMR measures total action disagreement; FIR and MIR separate unnecessary and missed interventions.

\subsection{Outcome Gaps}
The physical outcomes are CO$_2$ exceedance fraction above 1000 ppm, mean and peak latent CO$_2$, mean and total energy, mean comfort, and intervention duty cycle. For each non-oracle policy, paired outcome gaps are
\begin{align}
\Delta E_{\mathrm{CO2}} &= E_{\mathrm{CO2},p}-E_{\mathrm{CO2},o},\\
\Delta E_{\mathrm{energy}} &= E_{\mathrm{energy},p}-E_{\mathrm{energy},o},\\
\Delta E_{\mathrm{comfort}} &= E_{\mathrm{comfort},o}-E_{\mathrm{comfort},p},
\end{align}
where a positive comfort gap denotes worse comfort than the oracle. Reported intervals are empirical 2.5th--97.5th percentiles across paired simulation runs and are termed \emph{95\% Monte Carlo intervals}; they are not confidence intervals for a real population.

\subsection{Ground-Truth Regret}
An optional scalar objective is included to study policy-ranking sensitivity. Define
\begin{align}
C_{\mathrm{air}} &= E_{\mathrm{CO2}},\\
C_{\mathrm{energy}} &= \frac{E_{\mathrm{total\text{-}energy},p}}{E_{\mathrm{total\text{-}energy},P0}}-1,\\
C_{\mathrm{discomfort}} &= \frac{100-E_{\mathrm{comfort}}}{100},
\end{align}
and
\begin{equation}
J_{\mathbf w}=w_a C_{\mathrm{air}}+w_e C_{\mathrm{energy}}+w_c C_{\mathrm{discomfort}}.
\end{equation}
Ground-Truth Regret (GTR) is defined as the signed oracle-relative utility gap
\begin{equation}
\mathrm{GTR}_{\mathbf w}(p)=J_{\mathbf w}(p)-J_{\mathbf w}(o).
\end{equation}
Five weight profiles are evaluated: balanced $(1/3,1/3,1/3)$, air-quality priority $(0.6,0.2,0.2)$, energy priority $(0.2,0.6,0.2)$, comfort priority $(0.2,0.2,0.6)$, and safety-heavy $(0.7,0.15,0.15)$. Because P1 is an information-reference policy rather than an optimizer of $J_{\mathbf w}$, this signed quantity can be negative and should not be interpreted as classical nonnegative regret. GTR is therefore secondary to the separate decision and outcome metrics.

\subsection{Stress Scenarios}
The experiments are:
\begin{enumerate}[leftmargin=*]
    \item \textbf{Nominal benchmark:} 100 paired runs, nominal drift, 0.8\% explicit missingness.
    \item \textbf{Drift sweep:} severity $\{0,0.5,1,2,4,8\}\times$ nominal, with 1\% missingness.
    \item \textbf{Missingness sweep:} 0--20\% missing observations at nominal drift.
    \item \textbf{Combined grid:} drift $\{0,1,2,4\}\times$ crossed with missingness $\{0,0.05,0.10,0.20\}$.
    \item \textbf{Distribution shift:} occupancy $+25\%$, outdoor temperature $+5^{\circ}$C, filter efficiency multiplied by 0.85, and outdoor PM$_{2.5}$ variability multiplied by 1.75, all applied simultaneously.
    \item \textbf{Fault-aware ablation:} raw; rolling filter only; missing-data fallback only; jump/range rejection only; and the full method, under nominal sensing and under drift $4\times$ plus 20\% missingness.
    \item \textbf{GTR sensitivity:} post-hoc ranking under the five utility profiles.
\end{enumerate}

The stress magnitudes above are controlled scenario-design choices rather than estimates of real-world fault probabilities or operating frequencies. In the distribution-shift experiment, occupancy and outdoor temperature directly affect the reported closed-loop CO$_2$, energy, and comfort pathways. Filter efficiency and outdoor PM$_{2.5}$ are retained as auxiliary context variables and do not define the CO$_2$ control decision.

For the combined grid, material-divergence screening is applied to the three sensor-driven policies P2--P4 across the 16 grid cells, giving 48 policy--condition combinations. \emph{Material divergence} is predeclared as any of: DMR $\geq0.10$, $|\Delta E_{\mathrm{CO2}}|\geq0.02$, or $\Delta E_{\mathrm{comfort}}\geq1.0$ point. These thresholds are analysis choices, not universal safety thresholds.

\section{Results}
\label{sec:results}

\subsection{Nominal Closed-Loop Behavior}
Table~\ref{tab:nominal} reports the nominal means. No control yields a CO$_2$ exceedance fraction of 0.482 and mean latent CO$_2$ of 1416 ppm. The oracle reduces exceedance to 0.413 and increases mean energy from 3.188 to 3.588 kWh/interval while improving comfort from 73.29 to 84.34. This is the simulated air-quality--energy trade-off built into the model.

The raw-sensor policy almost matches oracle-level outcomes despite nonzero action disagreement. Its DMR is 0.0259, split nearly evenly between FIR=0.0131 and MIR=0.0129. Yet its mean CO$_2$ exceedance difference from the oracle is only $-0.000125$ with a 95\% Monte Carlo interval $[-0.00465,0.00469]$; mean energy differs by 0.000218 kWh/interval with interval $[-0.00297,0.00296]$; and the comfort gap is 0.0011 points with interval $[-0.030,0.038]$.

\begin{table*}[t]
\caption{Nominal policy benchmark (100 paired Monte Carlo runs). Means are shown; DMR/FIR/MIR are fractions of decision steps.}
\label{tab:nominal}
\centering
\small
\begin{tabular}{lrrrrrrrr}
\toprule
Policy & CO$_2$ exceed. & Mean CO$_2$ & Energy & Comfort & Duty & DMR & FIR & MIR \\
 & fraction & (ppm) & (kWh/int.) & (0--100) & cycle & & & \\
\midrule
No control & 0.4823 & 1416.13 & 3.1881 & 73.290 & 0.0000 & 0.4128 & 0.0000 & 0.4128 \\
Oracle & 0.4127 & 1036.77 & 3.5876 & 84.340 & 0.4127 & 0 & 0 & 0 \\
Raw sensor & 0.4126 & 1036.79 & 3.5879 & 84.338 & 0.4129 & 0.0259 & 0.0131 & 0.0129 \\
Filtered & 0.4119 & 1039.78 & 3.5864 & 84.233 & 0.4114 & 0.0631 & 0.0309 & 0.0322 \\
Fault-aware & 0.4119 & 1039.78 & 3.5864 & 84.233 & 0.4114 & 0.0631 & 0.0309 & 0.0322 \\
\bottomrule
\end{tabular}
\end{table*}

\begin{figure*}[t]
\centering
\includegraphics[width=0.95\textwidth]{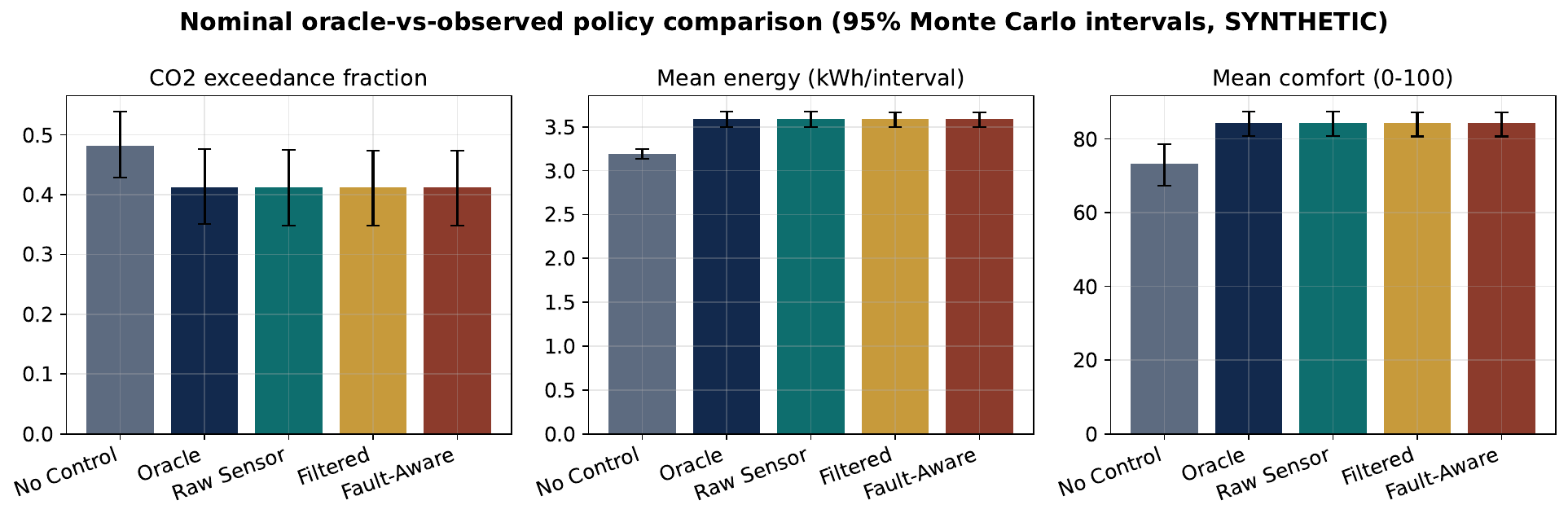}
\caption{Nominal physical outcomes with 95\% Monte Carlo intervals. The practical policies are close to the oracle in aggregate CO$_2$, energy, and comfort even though their decision sequences are not identical.}
\label{fig:outcomes}
\end{figure*}

\begin{figure}[t]
\centering
\includegraphics[width=\linewidth]{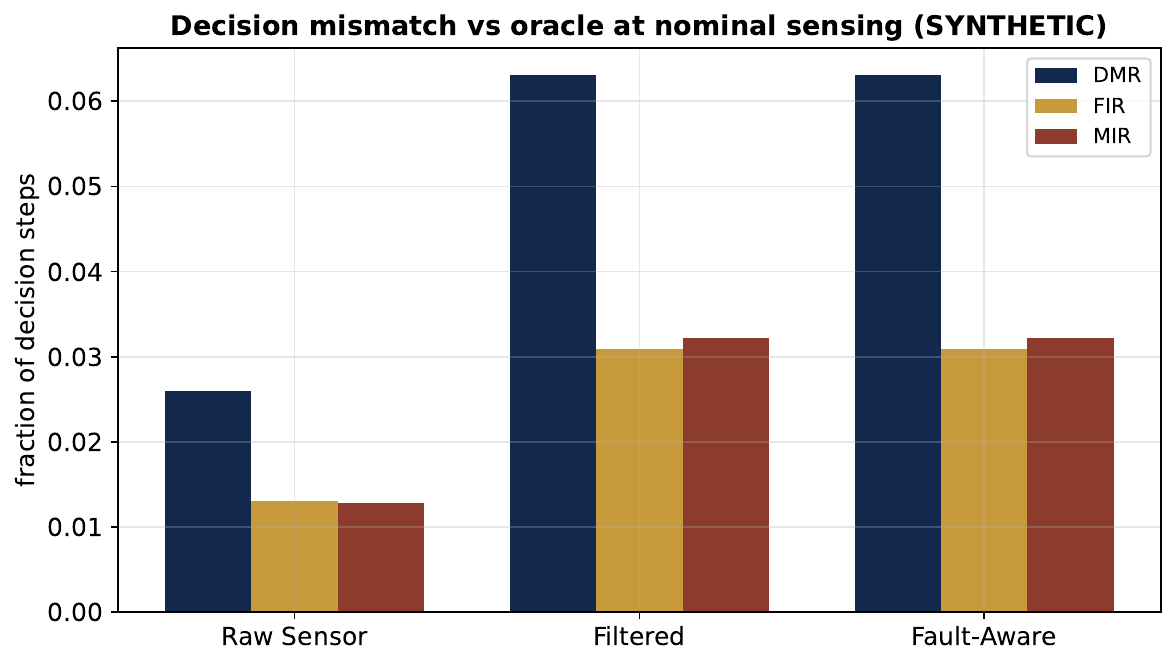}
\caption{Nominal decision mismatch relative to the oracle. A three-sample rolling median increases DMR, FIR, and MIR compared with the raw-sensor controller.}
\label{fig:dmr}
\end{figure}

The filtered and fault-aware policies are also close to the oracle in aggregate outcomes (Fig.~\ref{fig:outcomes}), but they disagree more often with the oracle. Both have DMR=0.0631, roughly 2.4 times the raw-sensor DMR (Fig.~\ref{fig:dmr}). Their mean comfort is 0.107 points lower than the oracle, with a 95\% Monte Carlo interval of $[0.066,0.154]$ for the gap. Although small in absolute magnitude, the direction is consistent across the paired simulation runs.

\subsection{Drift Stress: Disagreement Rises Faster Than Outcome Error}
Increasing drift produces the expected increase in action disagreement, especially for the raw controller. With zero drift, raw DMR is 0.0121; at nominal drift it is 0.0260; at $4\times$ drift it reaches 0.0445; and at $8\times$ drift it reaches 0.0700. At $8\times$ drift, FIR=0.0407 and MIR=0.0294.

\begin{figure*}[t]
\centering
\begin{minipage}{0.49\textwidth}
\centering
\includegraphics[width=\linewidth]{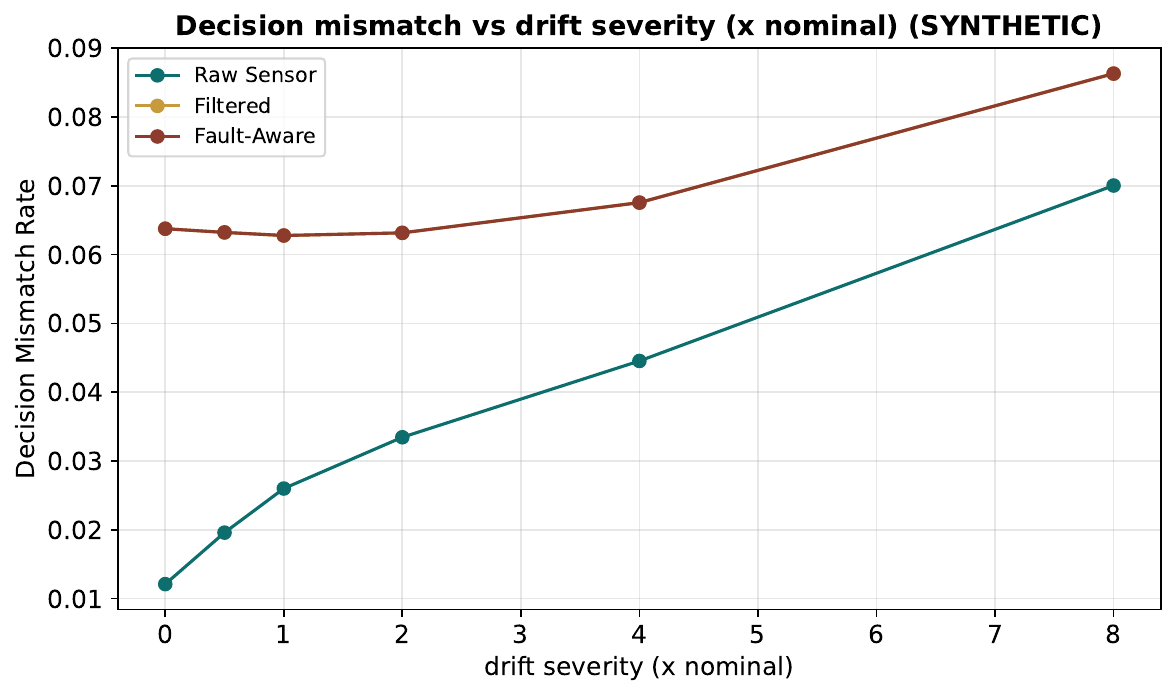}
\end{minipage}\hfill
\begin{minipage}{0.49\textwidth}
\centering
\includegraphics[width=\linewidth]{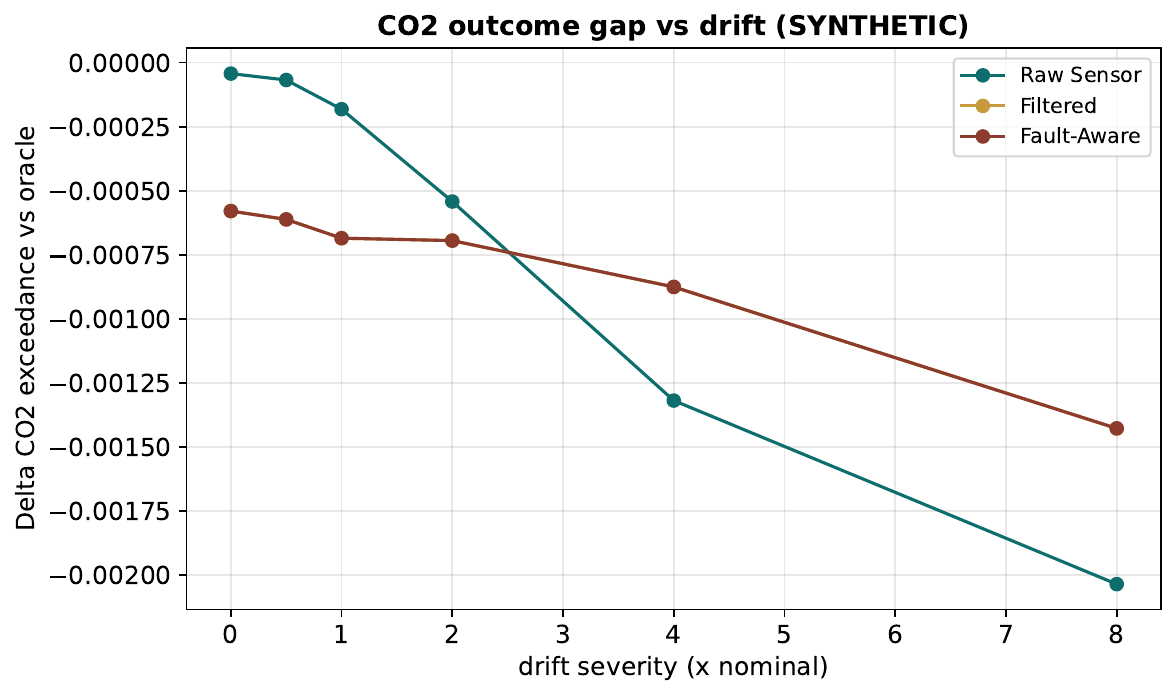}
\end{minipage}
\caption{Effect of increasing drift severity. Left: decision mismatch grows with drift. Right: the CO$_2$ exceedance gap relative to the oracle remains small over the same range.}
\label{fig:drift}
\end{figure*}

Despite this larger decision mismatch, the raw controller's physical gaps remain modest even at the strongest drift level: $\Delta E_{\mathrm{CO2}}=-0.00204$, $\Delta E_{\mathrm{energy}}=+0.01093$ kWh/interval, and $\Delta E_{\mathrm{comfort}}=0.0756$ points. The filtered/fault-aware policies reach DMR $\approx0.0863$ at $8\times$ drift with comfort gap $\approx0.183$ points. Figure~\ref{fig:drift} illustrates this different scale of growth: decision mismatch becomes visible long before the physical outcome gaps become large.

\subsection{Missing Observations and Combined Stress}
With 20\% missing observations at nominal drift, the raw controller reaches DMR=0.0389, FIR=0.0195, and MIR=0.0193. Its CO$_2$ exceedance gap is $-0.000316$, energy gap is $+0.000200$ kWh/interval, and comfort gap is 0.0343 points. At the same missingness level, the filtered and fault-aware policies have DMR $\approx0.0688$ and comfort gap $\approx0.177$ points.

\begin{figure}[t]
\centering
\includegraphics[width=\linewidth]{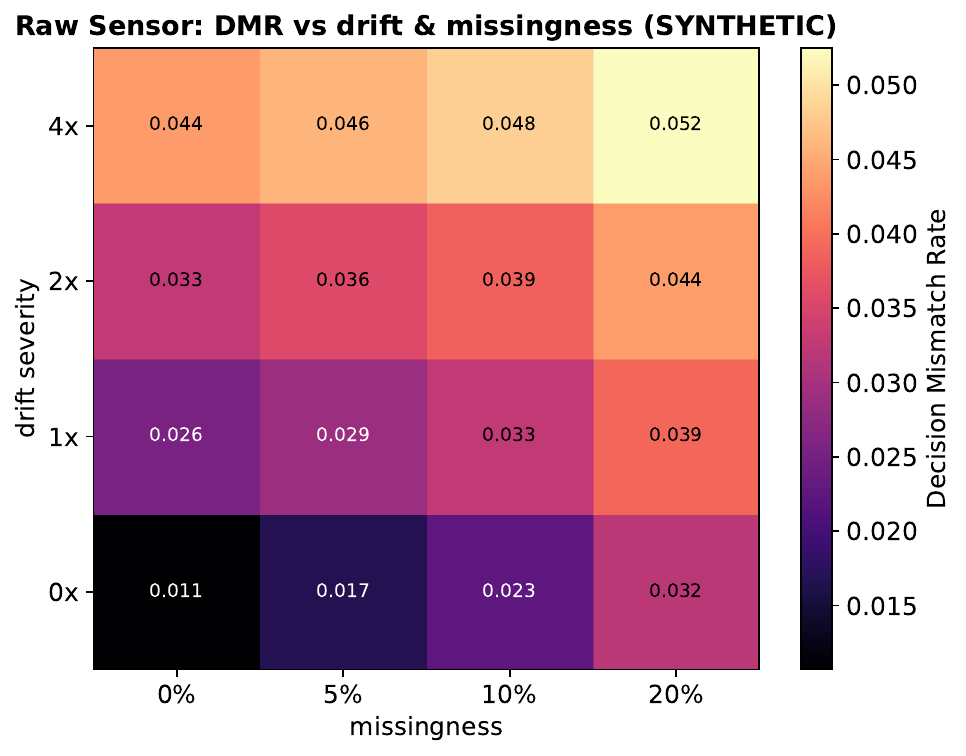}
\caption{Raw-sensor decision mismatch over the combined drift--missingness grid. DMR rises monotonically with stress but remains below the predeclared 0.10 materiality threshold in the evaluated region.}
\label{fig:failuremap}
\end{figure}

The two-dimensional grid combines drift up to $4\times$ with missingness up to 20\%. At the most severe grid point, raw DMR is 0.0525 and comfort gap is 0.0559 points, whereas the filtered and fault-aware DMR values are approximately 0.0735 with comfort gaps of approximately 0.206 points. None of the 48 sensor-driven policy--condition combinations crosses the predeclared material-divergence criteria. Figure~\ref{fig:failuremap} shows the raw-sensor DMR surface.

The absence of threshold crossings should not be interpreted as a general robustness guarantee. It is a property of the declared model, controller, stress ranges, and materiality choices. Nevertheless, it directly supports the paper's central point: within this experiment, nonzero sensor-induced action disagreement does not automatically imply material aggregate outcome degradation.

\subsection{Unseen Operating Distribution}
The distribution-shift scenario simultaneously increases occupancy by 25\%, raises outdoor temperature by $5^{\circ}$C, reduces filter efficiency by 15\%, and increases outdoor PM$_{2.5}$ variability by 75\%. Under this shift, the oracle's CO$_2$ exceedance rises from 0.413 to 0.464 and mean comfort decreases from 84.34 to 80.41. For the reported closed-loop outcomes, the harder regime is driven primarily by the occupancy and temperature perturbations.

\begin{figure*}[t]
\centering
\includegraphics[width=\linewidth]{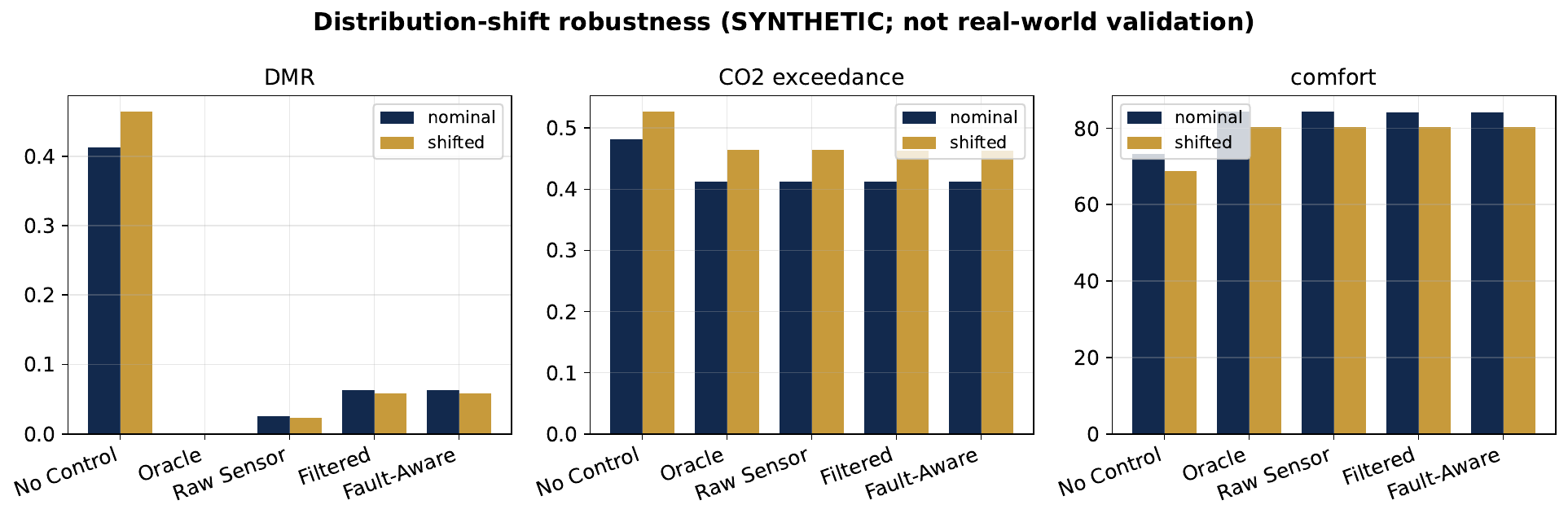}
\caption{Nominal versus shifted operating environment. The shift changes the physical operating regime, but the raw controller remains close to the oracle on the reported aggregate outcomes.}
\label{fig:shift}
\end{figure*}

The raw controller nevertheless remains close to the oracle: shifted DMR is 0.0228, CO$_2$ exceedance is 0.46381 versus 0.46383 for the oracle, mean energy is 3.41318 versus 3.41311 kWh/interval, and mean comfort is 80.403 versus 80.406. Filtered and fault-aware policies again have higher DMR (0.0581) while retaining similar aggregate outcomes. Figure~\ref{fig:shift} summarizes nominal-to-shift behavior.

\subsection{Fault-Aware Ablation: Added Filtering Can Hurt Agreement}
The ablation result is an important negative finding. At nominal sensing, the raw controller (A0) has DMR=0.0259. Adding only the missing-data occupancy fallback (A2) leaves DMR unchanged. Adding only jump/range rejection (A3) also leaves DMR essentially unchanged at 0.0260. In contrast, the rolling-median variant (A1) raises DMR to 0.0631; the full fault-aware method (A4) has the same DMR because the filter dominates its behavior.

\begin{figure}[t]
\centering
\includegraphics[width=\linewidth]{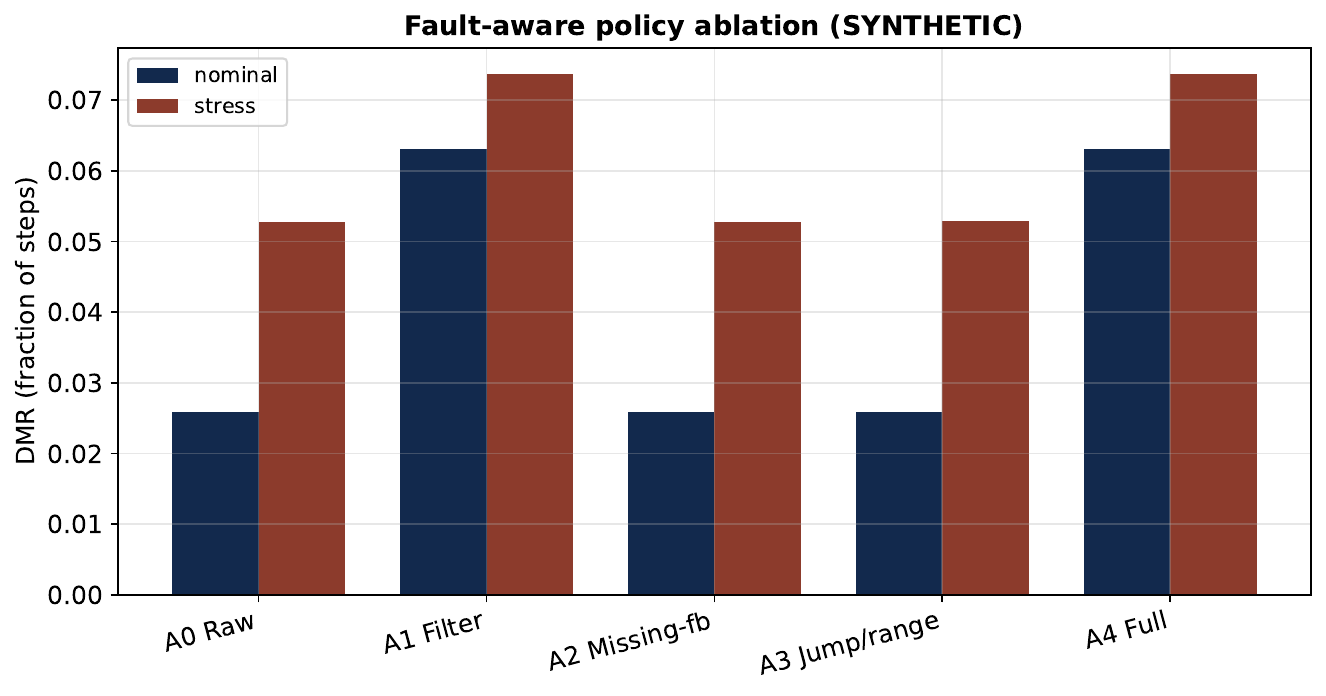}
\caption{Fault-aware ablation under nominal and stressed sensing. The rolling median is the main source of increased decision mismatch in the evaluated controller.}
\label{fig:ablation}
\end{figure}

Under the stress setting of $4\times$ drift and 20\% missingness, A0 has DMR=0.0527, A3 has 0.0529, and A4 has 0.0737. The corresponding comfort gaps are 0.0482, 0.0547, and 0.1932 points, respectively. Thus the added temporal smoothing does not improve oracle agreement under these dynamics; it introduces lag around the 1000 ppm threshold and produces extra false and missed interventions. Figure~\ref{fig:ablation} visualizes the ablation.

\subsection{Objective-Dependent Ground-Truth Regret}
\begin{figure}[t]
\centering
\includegraphics[width=\linewidth]{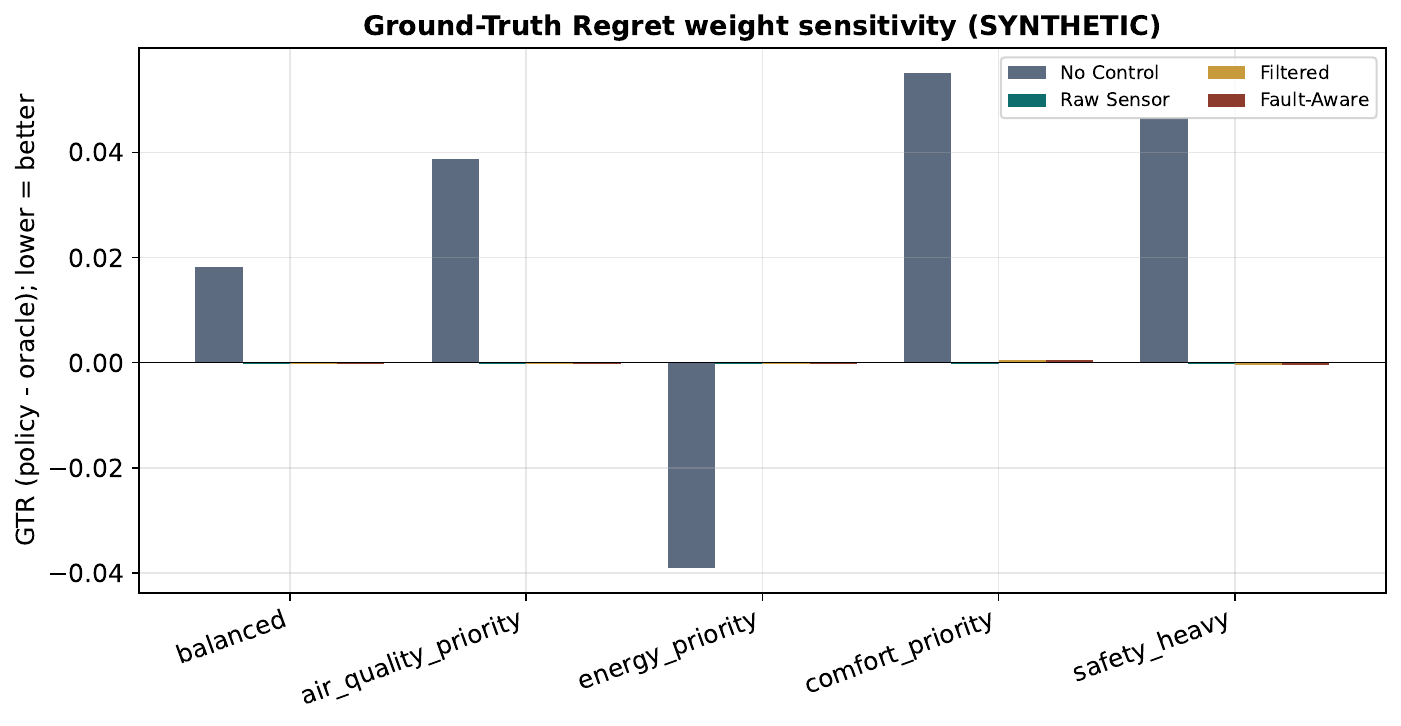}
\caption{Ground-Truth Regret sensitivity to utility weights. Policy ranking changes with the objective profile, so GTR is reported as a secondary, preference-dependent summary rather than a universal score.}
\label{fig:gtr}
\end{figure}

The GTR analysis illustrates why scalar policy rankings must be interpreted together with the chosen utility. Among non-oracle policies, the filtered controller ranks best under balanced, air-quality-priority, and safety-heavy weights; no control ranks best under energy-priority weights; and the raw controller ranks best under comfort-priority weights. The ranking therefore changes across profiles (Fig.~\ref{fig:gtr}). This is not a contradiction: the policies trade air quality, energy, and comfort differently, and scalarization exposes the preference assumptions required to call one policy ``best.''

\section{Discussion}
\label{sec:discussion}

\subsection{Decision Error Is Not the Same as Outcome Error}
The most important result is the separation between decision disagreement and physical consequence. Under nominal sensing, the raw controller disagrees with the oracle on roughly one in 39 decision points, but its aggregate CO$_2$, energy, and comfort outcomes are nearly indistinguishable from the oracle. Even as drift and missingness increase, DMR grows much more visibly than the outcome gaps.

Several mechanisms explain this behavior within the model. First, false interventions and missed interventions are relatively balanced for the raw policy, so some action differences cancel in aggregate duty cycle. Second, the first-order CO$_2$ dynamics smooth short-lived action differences; a one-step mismatch does not instantaneously reset the physical state. Third, the controller is threshold-based, so many disagreements occur near the threshold where adjacent actions may have only limited long-horizon effect. Finally, ventilation actions are bounded and the same exogenous trajectories are shared across policies.

This observation has practical methodological implications. A sensor-quality study that reports only measurement error may overstate or understate operational significance. Conversely, a control study that reports only aggregate outcomes may hide substantial action instability. Digital-twin robustness evaluation should therefore report at least three layers together: observation error, decision disagreement, and outcome degradation.

\subsection{More Preprocessing Is Not Automatically More Robust}
The ablation study cautions against assuming that a more elaborate sensor-processing pipeline is necessarily safer. Here, a three-sample rolling median adds approximately 30 minutes of effective temporal memory around a 15-minute control cadence. That lag increases disagreement with the oracle around threshold transitions. The full fault-aware controller therefore performs worse than the raw controller on DMR, even though its physical outcomes remain close to the oracle.

This is a useful negative result rather than a failed method. Sensor filtering is often motivated by noise suppression, but a filter can introduce phase delay or persistence that is undesirable for threshold-triggered control. Robustness must be evaluated against the downstream decision objective, not only against signal smoothness.

\subsection{Verification Versus Validation}
The framework passed automated checks for information separation, one-step timing, missing-data semantics, CRN alignment, and metric correctness. Such checks establish implementation verification: the program behaves according to the declared experimental design. They do not establish that the simulator is a valid representation of a real building. Simulation-model verification and operational validation are distinct activities \cite{sargent2010vv}.

This distinction is especially important here because the absolute physical parameters are intentionally synthetic. The plausibility audit found latent CO$_2$ values with a median near 893 ppm but a 99th percentile above 5600 ppm and a maximum above 8000 ppm in the base simulator. The auxiliary latent PM$_{2.5}$ model can also produce small negative values because additive noise is not clipped at zero. These facts do not invalidate the closed-loop software experiment, but they preclude claims that the absolute distributions represent a specific physical building. Accordingly, the present paper interprets relative controller behavior inside the model, not real-building air-quality or energy performance.

\section{Limitations and Future Work}
\label{sec:limitations}

First, the digital twin is synthetic and uncalibrated. Real deployment would require parameter identification, calibration, and validation against measured building data. Public faulted/fault-free HVAC datasets such as \cite{granderson2023dataset} provide one route for future sensor-layer validation, while operational building studies such as \cite{koo2026synchronization} illustrate the additional work required for in-situ synchronization.

Second, the control law is intentionally simple: a single CO$_2$ threshold with a fixed ventilation multiplier. More complex systems may show stronger propagation from decision errors to outcomes. Model-predictive control, multi-zone coupling, actuator constraints, minimum on/off times, and economic objectives could all change the observed resilience.

Third, the paper-specific CO$_2$ dynamics use a 45-minute first-order time constant. This choice improves closed-loop realism relative to instantaneous steady-state response but has not been calibrated or subjected to a dedicated time-constant sensitivity study. Future work should vary $\tau$ and examine whether decision/outcome decoupling persists across faster and slower physical dynamics.

Fourth, the drift process is a zero-mean random walk scaled by severity. Persistent calibration bias may be more damaging to a threshold controller because it can systematically move observations across the control boundary. A targeted additive-bias sweep is therefore an important next experiment.

Fifth, the materiality thresholds used in the combined grid are declared analytical choices, not safety standards. Different applications may require much tighter tolerances. Similarly, GTR depends on subjective utility weights and should not replace disaggregated outcome reporting.

Finally, the fault-aware controller is not presented as a new optimal method. Its ablation result is deliberately reported even though it underperforms the raw controller on decision agreement. Future work should design fault-aware policies that account explicitly for temporal delay, uncertainty, and actuator dynamics rather than applying smoothing heuristics alone.

\section{Conclusion}
\label{sec:conclusion}

This paper presented a ground-truth-aware closed-loop simulation framework for evaluating how sensor corruption propagates from measurement error to control decisions and physical outcomes in a synthetic smart-building digital twin. The framework separates latent CO$_2$ state from observed sensor state, enforces a one-step information delay, compares practical controllers with an oracle under common random numbers, and reports both decision mismatch and outcome gaps.

Across nominal sensing, drift sweeps, missingness sweeps, a combined stress grid, and an unseen operating distribution, sensor corruption increased decision disagreement substantially more than it changed aggregate CO$_2$, energy, or comfort outcomes. The raw-sensor controller had 2.59\% nominal DMR yet essentially oracle-equivalent aggregate outcomes; at $8\times$ nominal drift its DMR rose to 7.00\% while physical gaps remained small. None of the 48 sensor-driven policy--condition combinations in the combined stress grid crossed the study's predeclared material-divergence thresholds. The ablation further showed that rolling-median filtering increased mismatch and that the full fault-aware controller did not outperform the raw controller on decision agreement. GTR rankings changed with utility weights, underscoring the importance of reporting multi-objective trade-offs explicitly.

The main methodological conclusion is that \emph{sensor error, decision error, and outcome error are related but distinct}. Digital-twin evaluation should measure all three rather than assuming that better sensing, smoother signals, or closer action agreement automatically implies better physical performance. The present results are deliberately limited to a verified synthetic model; real-building calibration and validation are required before operational claims can be made.

\section*{Data and Code Availability}
All data used in this study are fully synthetic. The supplementary experiment package accompanying this manuscript contains the closed-loop source code, configuration files, automated tests, full Monte Carlo result tables, publication figures, and machine-readable result summaries.

\section*{Declaration of Generative AI and AI-Assisted Technologies in the Manuscript Preparation Process}
During the preparation of this work, the author used OpenAI ChatGPT to improve the English language and presentation. 

\bibliographystyle{elsarticle-num}
\bibliography{references}

\end{document}